\documentclass[%
reprint,
 amsmath,amssymb,
 aps,
 prl,
]{revtex4-2}
\usepackage{natbib}
\usepackage{comment}
\usepackage{graphicx}
\usepackage{dcolumn}
\usepackage{bm}

\usepackage [english]{babel}
\usepackage [autostyle, english = american]{csquotes}
\MakeOuterQuote{"}

\begin{document}

\preprint{APS/123-QED}

\title{A Toy Model of Quantum Measurement with Experimentally Falsifiable Predictions}

\author{Aaron Santos}
 \affiliation{DNP123 Company, Indianola, IA 50125}


\date{\today}

\begin{abstract}
I describe a toy model of quantum measurement in which wave function collapse is described as a stochastic entropically-driven event guided by interactions between a measured two-state particle and an Ising-like measurement device. The model reproduces existing experimental results and suggests new calorimetric experiments that can be used to test hidden variables theories. When applied to entangled particles, the model predicts energy cannot be transferred faster than the speed of light, but that certain schemas would allow for nonlocal information transfer.  
\end{abstract}

\maketitle


The Copenhagen interpretation of quantum mechanics states that the wave function evolves deterministically via the Schrödinger equation until a measurement occurs, at which point the wave function instantaneously and non-deterministically transitions to an eigenstate of the measurement operator. Einstein, Podolsky, and Rosen challenged this interpretation, arguing that the locality required by special relativity suggests there must be hidden variables determining the outcome of an experiment prior to measurement.\cite{einstein1935can} Bell derived an inequality that local hidden variables must satisfy\cite{bell1964einstein,bell2004speakable}; however, this inequality is violated by certain experiments involving entangled particles.\cite{aspect1982experimental,weihs1998violation,rowe2001experimental,groblacher2007experimental,handsteiner2017cosmic} These tests of Bell's inequality rule out some versions of local hidden variables, specifically those that maintain the statistical independence of the measurement devices. Given these facts, researchers have developed competing theories and interpretations of quantum mechanics. Pilot wave and objective collapse theories abandon locality,\cite{bohm1952suggested,ghirardi1986unified,penrose1996gravity} while superdeterminism abandons measurement independence.\cite{brans1988bell,hall2010local,donadi2022toy} The many-worlds interpretation circumvents the issue by introducing multiple realities.\cite{everett1957relative} Testing these theories experimentally remains a challenge. In this paper, I propose a toy model of quantum measurement that explicitly considers interactions with individual particles in the measurement device. The model offers predictions that can be used to experimentally test certain classes of hidden variable theories.


The theories mentioned above presuppose that quantum measurement necessitates either alternative dynamics from Schrödinger's equation or an interpretation vastly different from the Copenhagen interpretation. In this paper, I posit that the time-dependent Schrödinger equation is adequate and explore the question, "What would happen to a wave function in a superposition state if it were to interact with a measurement device composed of particles undergoing random thermal motions?" In this scenario, the measurement device acts as a time-varying external potential on the measured quantum particle. The wave function's collapse to a single eigenstate of the measurement operator occurs through the natural time evolution of Schrödinger's equation as dictated by the time-dependent Hamiltonian, $H(t)$. While we can abstract the measurement device as an external potential for a system consisting solely of the quantum particle, the Hamiltonian of a system that includes both the measured quantum particle and the measurement device must include (1) the interaction energy between the measured quantum particle and the particles within the measurement device, and (2) the interactions among the particles inside the device. The interaction energy between a quantum particle and a macroscopic measurement device is inherently stochastic due to the random motion of particles within and surrounding the device, introducing noise into the Hamiltonian. This noise, varying rapidly over time, influences the wave function's time evolution, stochastically and irreversibly steering it into one of the measurement eigenstates. As a macroscopic system, the measurement device is assumed to adhere to classical Boltzmann statistics, where each state's probability is proportional to a Boltzmann factor $P \propto \exp(-\beta H)$, with $\beta = 1/T$ representing the inverse temperature.

A measurement device must possess the following properties: (1) It should contain a finite number of internal particles that interact with the quantum particle. (2) Each macroscopic state of the measurement device (i.e., the unmeasured state and each of the possible outcomes that correspond to eigenstates of the measurement operator) must correspond to distinct microscopic configurations of the internal particles. (3) The measurement device must start in an unmeasured state and, solely through interactions of its internal particles with the quantum particle, evolve into one of the possible measurement outcomes. Lastly, (4) the statistics of the measurement should reflect the probabilities defined by the initial state of the quantum particle being measured.

In this toy model, I consider the measurement of a two-state particle. Generally, the state of the particle is represented as a superposition:

\begin{equation}
\psi = c_A \left|A\right> + c_B \left| B \right>,
\end{equation}
where, $A$ and $B$ denote the two eigenstates of the measured observable's operator. The coefficients $c_A$ and $c_B$ define the probabilities of states $A$ and $B$ as $P(A)=| c_A|^2$ and $P(B)=|c_B|^2$, respectively. The normalization condition requires that $c_A^2+|c_B|^2$.

To model the internal state of the measurement device, I employ a $L\times L$ two-dimensional Ising-like model on a square lattice. This approach is chosen for its simplicity and because it meets the previously outlined criteria. In this model, each spin $\sigma_i$ represents the state of particle $i$ within the measurement device. Spins that are nearest neighbors interact with a specific energy,
\begin{equation}
H_{internal} = -J \sum_{<ij>} \sigma_i*\sigma_j
\end{equation}
The macroscopic state of the measurement device is defined by the magnetization $M=\sum\sigma_i/L^2$, where a net positive magnetization corresponds to the particle being measured in state $A$, and a net negative magnetization corresponds to state $B$. In this context, I use "spin" and "magnetization" in alignment with the familiar terminology of the Ising model, but these terms need not imply spin and magnetization in the conventional sense. They serve as placeholders representing, respectively, the state of an individual particle within the measurement device and the device's macroscopic state. For instance, "magnetization" could signify the velocity at which a macroscopic indicator needle is deflected in the device, while "spin" might represent the direction of velocity (left or right) of individual particles in the needle. In this framework, the unmagnetized state, signifying the measurement device's state prior to measurement, would be evenly split between up and down spin states. This split represents the random thermal motion of particles in the needle, whereas a net magnetization would indicate a needle deflecting to the left (particle in state $A$) or right (particle in state $B$). Similar to how the traditional Ising model provides qualitatively correct behavior but does not accurately predict the critical temperature or magnetization versus temperature curve for any real ferromagnet, this model is expected to yield only qualitatively correct predictions.

I model the interaction between the measurement device and the measured quantum particle as
\begin{equation}
H_{interaction} = -K\alpha\sum_i \sigma_i
\end{equation}
Here, $K$ represents the strength of interaction between the quantum particle and the spins within the measurement device. The parameter $\alpha\in [0,1]$ defines the state of the quantum particle. Specifically, $\alpha=1$ corresponds to the quantum particle being in eigenstate $A$; $\alpha=0$, to the particle being in eigenstate $B$; and $\alpha=0.5$, to the particle being in a 50/50 superposition between states $A$ and $B$. It is important to note that $\alpha$ does not necessarily equal the probability $P(A)$ of measuring the particle in state $A$. Rather, it is sufficient that $\alpha$ can be mapped to this probability.

I selected values for the model parameters based on the following reasoning.  Realistically, the quantum particle will only interact directly with particles in the measurement device over a relatively small range. The number $L^2$ of spins in the model corresponds to the number of particles the quantum particle interacts with, not the total number of particles in the measurement device. As such, the linear size $L$ of the measurement device must be greater than 1 but not macroscopically large. I considered values of $L$ ranging from 4 to 128. The temperature of the device must be high enough to prevent spontaneous magnetization. Given that the critical temperature of a two-dimensional Ising model with $J=1$ is $T_{crit}=2\ln (1+2)\approx2.269$, I chose $T\geq3$. Correspondingly, the interaction strength between the quantum particle and the measurement device, denoted by $K$, must be significant enough to overcome thermal fluctuations. For this reason, I chose $K=10$.

I employ the Metropolis Monte Carlo (MMC) algorithm to model the measurement process. It is important to recognize that the dynamics introduced by MMC are fictitious, as the algorithm is primarily designed to achieve correct equilibrium Boltzmann statistics. In the MMC algorithm, random trial moves are generated by selecting either a random spin from the measurement device or the quantum particle, each with equal probability. If a spin is chosen, it is flipped. Conversely, if the quantum particle is selected, its state is either increased or decreased by $\delta\alpha = 0.1$ with equal probability. The change in total energy, $\Delta H$, is then calculated, where the total energy is given by:
\begin{equation}
H=H_{internal}+H_{interaction}
\end{equation}
All moves resulting in a decrease in energy $\Delta H$ are accepted. Conversely, moves that increase the energy are accepted with a probability $P\propto \exp (-\beta \Delta H)$. To guarantee that the measurement device is in thermal equilibrium, I allow the device to equilibrate for 200 MMC steps per spin before introducing the quantum particle.

\begin{figure}[t]
\includegraphics[width=8.6cm]{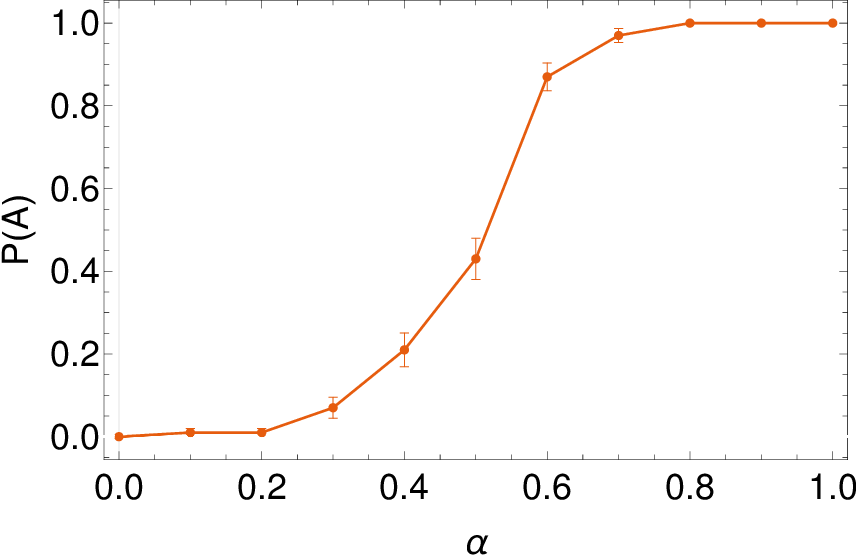}
\caption{\label{fig:epsart2} (color online) Plot of the measured probability $P(A)$ of finding the particle in state $A$ versus the initial value of $\alpha$. Since the plot is a monotonically increasing function, $\alpha$ can be mapped to $P(A)$ indicating that the model reproduces the results of quantum mechanics. Taken for $L=4$, $J=1$, $T=3$, and $K=10$.
}
\label{fig:reproduceQM}
\end{figure}

In all simulation runs, the system evolves either to $\alpha=1$ and $M=1$ or to $\alpha=0$ and $M=-1$. This outcome is essential to mirror the result of a real quantum measurement, where both the quantum particle and the measurement device end in an eigenstate of the observable operator. To validate that the model reproduces the probabilities dictated by the quantum particle's initial state, I conducted 100 simulation runs for varying $\alpha$ values, ranging from 0 to 1 in increments of 0.1. Subsequently, I calculated the probability $P(A)$ of the particle being measured in state $A$ for each $\alpha$ value. As demonstrated in the plot in Fig. \ref{fig:reproduceQM}, the measured probability $P(A)$ is a monotonically increasing function of $\alpha$. This trend indicates that $\alpha$ can be mapped to $P(A)$, affirming that the model accurately replicates the probabilities predicted by quantum mechanics.

\begin{figure}[t]
\includegraphics[width=8.6cm]{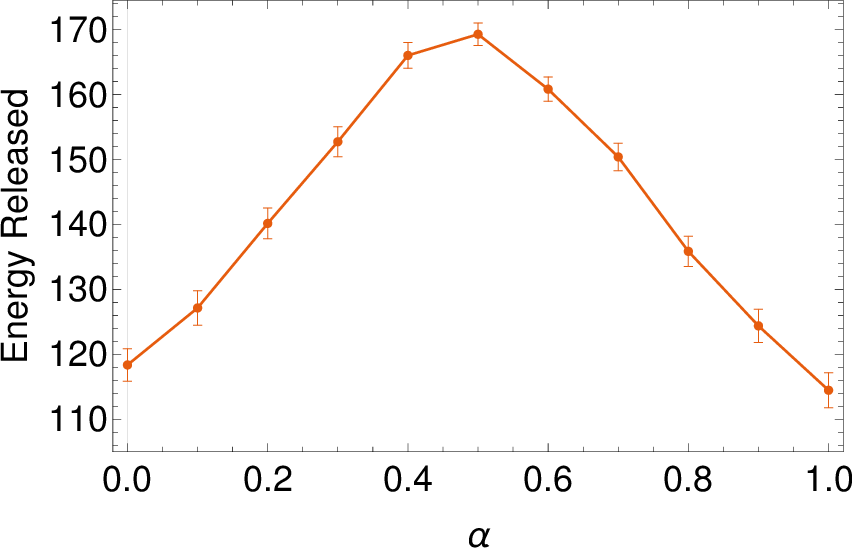}
\caption{\label{fig:epsart2} Plot of the heat released by the measurement device to its surroundings as a function of $\alpha$ for $L=4$, $J=1$, $T=3$, and $K=10$.
}
\label{fig:heatVSalpha}
\end{figure}

For a model to be useful, it is critical that it makes predictions that can be tested experimentally. Central to the model described here is the exchange of energy between the quantum particle and the measurement device, as well as between the measurement device and its surroundings in the form of heat. This scenario is similar to how a conventional thermometer operates, in that the measurement device inherently draws some energy from the system it measures. Calorimetric measurements of the heat exchange between the measurement device and its surroundings offer a method to test the model. To demonstrate how such an experiment might be conducted, I measured the heat dissipated from the measurement device as a function of $\alpha$ (see Fig. \ref{fig:heatVSalpha}). The results show that the energy released as heat is symmetric and peaks at $\alpha=0.5$, corresponding to the quantum particle initially being in a 50/50 superposition of states $A$ and $B$. The least heat transfer occurs when the quantum particle is in either eigenstate $A$ or $B$, but not in a superposition of both.

To empirically test this prediction, one could set up a stream of two-state quantum particles in either an eigenstate or a superposition state and measure the heat transferred out of the measurement device. Such an experiment would differentiate the current model from hidden variable theories where the quantum particle's state is predetermined before it enters the measurement device. The experimental outcomes could potentially disqualify certain classes of models. In particular, a discernible difference in heat released between particles in an eigenstate and those in a superposition would imply that stochastic processes within the measurement device dictate the observed eigenstate. This finding would challenge the validity of realism theories like pilot waves and superdeterminism, where experimental outcomes are pre-established. It would also undermine the rationale for the many-worlds interpretation, as wave function collapse would then resemble a stochastic process, akin to rolling dice, rather than an apparent break from the superposition suggested by the Schrödinger equation.

While the aforementioned experiment could be instrumental in testing certain classes of hidden variable theories, it does not address the question of whether the interaction between entangled quantum particles is local or nonlocal. Experimental violations of Bell’s inequality demonstrate that deterministic quantum mechanical theories cannot simultaneously be local and uphold the statistical independence of the measurement device. To explore this further while ensuring consistency with both relativity and Bell’s inequality, I assume that all interactions are local except those mediated between entangled particles. I simulated this by considering a pair of entangled particles 1 and 2 that are directed to separate measurement devices. The particles are prepared such that a measurement of particle 1 in state A necessitates particle 2 being measured in state B, and vice versa. The statistics of this scenario were again modeled using the MMC algorithm. In contrast to the original simulation where the lone quantum particle’s state was described by the parameter $\alpha$, in this setup, the states of particles 1 and 2 are determined by separate parameters $\alpha_1$ and $\alpha_2$, respectively. Because of their entanglement, the states of both particles can be described by a single parameter $\gamma$, where $\alpha_1=\gamma$ and $\alpha_2=1-\gamma$. Trial moves of $\gamma\rightarrow\gamma\pm\delta \gamma$ were generated randomly with equal probability with $\delta \gamma =0.1$. These moves were accepted or rejected based on a Boltzmann probability, similar to the previous approach. In this case, the Boltzmann probability accounts for the interactions of both particle 1 and 2 with their respective measurement devices,
\begin{equation}
P \propto \exp(-H_1/T_1)\exp(-H_2/T_2),
\end{equation}
where $H_1$ and $H_2$ are the total energies associated with measurement devices 1 and 2, and $T_1$ and $T_2$ are the temperatures of measurement devices 1 and 2, respectively.

\begin{figure}[t]
\includegraphics[width=8.6cm]{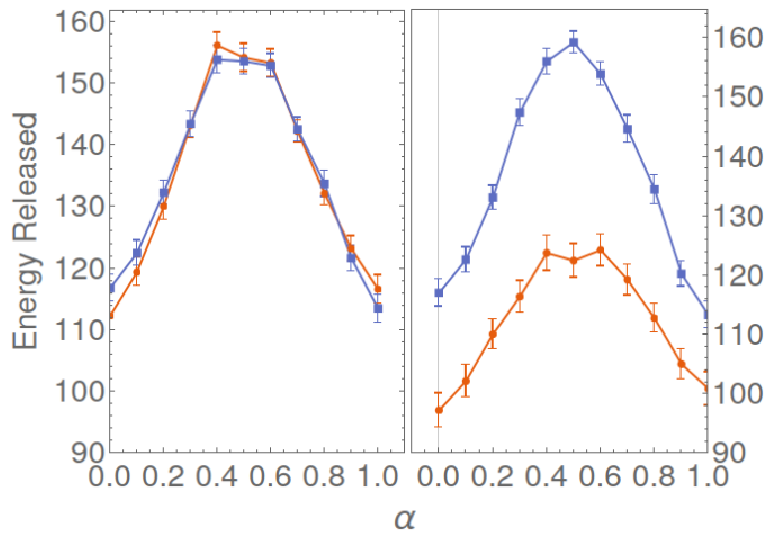}
\caption{\label{fig:epsart2} (color online) Plot of the heat released by the measurement device as a function of $\alpha$. (Left) Entangled particles are measured at devices 1 (blue square) and 2 (orange circle) with temperatures $T_1=T_2=3$. (Right) Entangled particles are measured at devices 1 (blue square) and 2 (orange circle) with temperatures $T_1=3$ and $T_2=6$. Taken for $L=4$, $J=1$, and $K=10$.}
\label{fig:heatVSalphaEntangle}
\end{figure}

The results of the experiment are illustrated in Fig. \ref{fig:heatVSalphaEntangle}. As expected by symmetry, when the measurement devices are at identical temperatures, they display the same $P(A)$ versus $\alpha$ curve. The local environment of each measurement device functions as a heat bath, facilitating energy exchange with the device. Raising the temperature of the heat bath connected to one of the devices results in a reduced amount of heat being transferred back to the local environment during measurement, as expected from classical thermodynamics. A question may arise regarding the entanglement’s nonlocal nature and whether it could enable energy transfer from the hotter to the colder measurement device. However, within the limits of the simulation, altering the temperature of measurement device 1 does not significantly impact the heat dissipation from measurement device 2. This outcome implies that energy cannot be transferred nonlocally, adhering to the speed of light limitations imposed by relativity.

However, if the model proves accurate, then there exists a measurable difference in heat transferred out of a measurement device for eigenstates and superposition states. This implies that while energy transfer is constrained by the speed of light, information might be communicated nonlocally through quantum entanglement. Consider the following communication scheme: Alice encodes a binary message where zeros and ones correspond to whether she has measured or not measured her quantum particle. Bob, on receiving each quantum particle, measures both the particle and the heat released by his measurement device. Given that more heat is emitted when the device measures a particle in a superposition state, Bob can infer whether Alice has measured her particle. Thus, Bob can decode messages from Alice without any energy being exchanged. Should such a phenomenon be experimentally verified, it would suggest that while energy transfer is restricted to the speed of light, information transfer might not be similarly limited. Since calorimetric measurements of this nature have not yet been performed, this hypothesis remains consistent with current experimental data on information transfer limits. If calorimetry experiments can distinguish whether particles are in eigenstates or superpositions, it would open the possibility of faster than light communication.

In summary, this paper introduced a toy model of quantum measurement that adheres to Schrödinger's equation. The model could be classified as a nonlocal hidden variables theory, with the hidden variables residing not in the wave function but within the measurement device. While aligning with current experimental evidence, the model also proposes novel predictions that allow for empirical testing against other quantum measurement theories. A key prediction is the discernible difference in heat emitted during a quantum measurement, depending on whether the particle is in an eigenstate or a superposition state. Consistent with the principles of relativity, the model posits that energy transfer is limited to the speed of light. However, it suggests that information transfer via entangled particles is not bound by this limitation. Experimental verification of the model could be pursued through calorimetric measurements of the measurement device, as heat would be released to the local environment increasing the total entropy of the universe. If validated, the model would imply that the collapse of a particle's wave function from a superposition to an eigenstate during measurement is an entropically-driven process, reducing the need for a fundamental revision of our understanding of quantum mechanics.

\bibliography{apssamp}
\bibliographystyle{apsrev4-2} 
\end{document}